\documentclass[a4paper,12pt]{article}
\usepackage{cite}
\usepackage{amssymb,amsmath,bm,xcolor}
\title{Analogue Hawking temperature of a laser-driven plasma}

\author{C. Fiedler \and D. A. Burton}

\begin{document}

\maketitle
Department of Physics, Lancaster University, Lancaster, LA1 4YB, UK
\begin{abstract}
We present a method for exploring analogue Hawking radiation using a laser pulse propagating through an underdense plasma. The propagating fields in the Hawking effect are local perturbations of the plasma density and laser amplitude. We derive the dependence of the resulting Hawking temperature on the dimensionless amplitude of the laser and the behaviour of the spot area of the laser at the analogue event horizon. We demonstrate one possible way of obtaining the analogue Hawking temperature in terms of the plasma wavelength, and our analysis shows that for a high intensity near-IR laser the analogue Hawking temperature is less than approximately $25$ K for a reasonable choice of parameters.
\end{abstract}

\section{Introduction}
Hawking radiation \cite{originalhawkingradiation} is a well established theoretical prediction that a black hole will evaporate, emitting a thermal spectrum of radiation in the process. The temperature of this radiation, called Hawking temperature, is inversely proportional to the mass of a black hole and thus it is very small, eluding experimental searches. However, it has been discovered that black holes are not the only phenomena that can generate Hawking radiation. Analogue gravity \cite{livingreview} investigates non-gravitational analogues of general relativity in various physical systems such as fluids \cite{originalUnruh, shallowtheory, generalwaterwave}, dielectric media \cite{dielectricnormal, dielectric2}, and Bose-Einstein condensates \cite{BEC}. In these systems, the field equations can be written as a massless wave equation on an effective background spacetime. Experiments performed in the context of analogue gravity include water waves \cite{watertankex1,watertankex2,watertankex3,watertankex4,watertankex5,watertankex6}, ultra-cold quantum gases \cite{gasex1,gasex2,gasex3,gasex4}, optics \cite{opticsex1,opticsex3,opticsex4,opticsex5,opticsex6}, and polaritons \cite{polaritonex}. Since experimental study of black holes is not feasible, analogue models provide the opportunity to test theoretical aspects of black hole physics in a laboratory environment.

The interest in analogue gravity is not only to reproduce gravitational phenomena in some analogue model, but to indirectly probe the simplifications that underpin calculations of black hole evaporation. Deviations of real analogue models from their simplest forms may shed light on the deviations that are likely to appear in the corresponding gravitational systems. See ref. \cite{livingreview} for a detailed discussion of the applicability of analogue models in this context. 

To our knowledge, the {\color{black}prospect} of using laser-driven plasma to study analogue black holes has not been thoroughly investigated. In this article we obtain the analogue Hawking temperature of a laser-driven underdense plasma. In the second section, we briefly describe how to obtain the non-linear field equations that underpin the work. In the third section we show that, under certain conditions, two effective Lorentzian metrics emerge when the field equations are linearised about an exact solution. In the fourth section we determine required properties of the fields in order to attain an effective metric analogous to the Schwarzschild metric, and we calculate the associated analogue Hawking temperature corresponding to using an intense near-IR laser. In the final section we summarise presented work.

\section{Laser-driven plasma}
\label{sec:laser-driven_plasma}
Quantities with tilde over them denote dimensionful variables which will be made dimensionless for convenience in the analysis. To first approximation, the properties of the plasma electrons can be described by a local energy-momentum relation and a continuity equation arising from charge conservation, whilst the laser pulse satisfies a local dispersion relation and conservation of wave action (i.e. conservation of classical photon number). This approach is underpinned by a separation of the laser-plasma dynamics into slowly evolving parts and rapidly oscillating parts that arises because the length scale of the internal oscillations of the laser pulse (i.e. the wavelength of the laser) is much shorter than the envelope of the laser pulse and the plasma wavelength. The rapidly oscillating quantities enter the slow dynamics via their local averages.

The relativistic energy-momentum relation
\begin{equation}
\label{energy-momentum_relation}
E^2 - c^2 {\bf p}^2 - c^2 e^2 \langle {\bf A}_0^2\rangle = m^2_e c^4 
\end{equation}
is satisfied by the averaged motion of the plasma electrons, where $E$ is the local relativistic energy of the averaged motion, ${\bf p}$ is the local relativistic kinetic momentum of the averaged motion and ${\bf A}_0$ is the vector potential of the laser pulse. The angle brackets in (\ref{energy-momentum_relation}) denote averaging over the fast internal oscillations of the laser pulse, and the third term on the left-hand side arises because $e {\bf A}_0$ can be identified with the momentum of fast oscillatory motion in the plane orthogonal to the direction of propagation of the laser pulse. The dispersion relation 
\begin{equation}
\label{local_dispersion_relation}
\omega_0^2 - c^2{\bf k}_0^2 = \omega_p^2
\end{equation}
is satisfied by the local angular frequency $\omega_0$ and the local wave vector ${\bf k}_0$ of the laser pulse. The local plasma frequency $\omega_p$ is given by
\begin{equation}
\omega_p = \sqrt{\frac{e^2 n}{\varepsilon_0 m_e \gamma}}
\end{equation}
where $\gamma$ is the local Lorentz factor of the averaged motion (i.e. $E=m_e c^2 \gamma$), and $n$ is the averaged number density, of the plasma electrons. Note that $E$, ${\bf p}$, $\langle {\bf A}_0^2\rangle$, $\omega_0$, ${\bf k}_0$, $\omega_p$, $n$, and $\gamma$ are fields.

The above physical quantities are coupled together by the conservation of wave action
\begin{equation}
\label{wave_action_conservation}
\partial_{\tilde t} (\langle {\bf A}_0^2\rangle\omega_0) + c^2\widetilde{\bm{\nabla}}\cdot(\langle {\bf A}_0^2\rangle{\bf k}_0) = 0
\end{equation}
and the continuity equation $\partial_{\tilde t} n + \widetilde{\bm{\nabla}}\cdot(n {\bf v}) = 0$, i.e.
\begin{equation}
\label{number_conservation}
\partial_{\tilde{t}} (\omega_p^2 E) + c^2\widetilde{\bm{\nabla}}\cdot(\omega_p^2 {\bf p}) = 0
\end{equation}
where ${\bf p} = m_e \gamma {\bf v}$. The symbol $\widetilde{\bm{\nabla}}$ denotes the gradient operator with respect to $\tilde{x}$, $\tilde{y}$, $\tilde{z}$. The system of field equations (\ref{energy-momentum_relation}), (\ref{local_dispersion_relation}), (\ref{wave_action_conservation}), (\ref{number_conservation}) was previously used in a self-consistent model \cite{envelopeequations} of the interaction between the laser pulse and the plasma wake it excites, in the context of electron acceleration.

If the forces due to the magnetic field sourced by the averaged current are negligible in comparison to the forces produced by the laser pulse then the flow of electrons can be chosen to be irrotational; thus, ${\bf p}$ can be expressed in terms of a potential $\widetilde\Psi$ as ${\bf p}=\widetilde{\bm{\nabla}}\widetilde\Psi$. Furthermore, if the electric field sourced by the averaged charge density can be neglected then one can determine $E$ using $E = -\partial_{\tilde t} \widetilde\Psi$. The expression
\begin{equation}
\label{average_A0_squared}
c^2 e^2 \langle {\bf A}_0^2\rangle = (\partial_{\tilde t}\widetilde\Psi)^2 - c^2(\widetilde{\bm{\nabla}}\widetilde\Psi)^2 - m^2_e c^4 
\end{equation}
for $\langle {\bf A}_0^2\rangle$ is then obtained from (\ref{energy-momentum_relation}). For completeness, it is straightforward to show that (\ref{energy-momentum_relation}) alongside the expressions $E=-\partial_{\tilde t}\widetilde{\Psi}$, ${\bf p}=\widetilde{\bm{\nabla}}\widetilde{\Psi}$, ${\bf v}={\bf p}/(m_e \gamma)$ solve the local equation of momentum balance
\begin{equation}
\label{local_momentum_balance}
\partial_{\tilde t} {\bf p} + ({\bf v}\cdot\widetilde{\bm{\nabla}}){\bf p} = -\frac{e^2}{2m_e\gamma}\widetilde{\bm{\nabla}}\langle {\bf A}_0^2\rangle
\end{equation}
where the force on the electrons is due entirely to the laser pulse. The term on the right-hand side of (\ref{local_momentum_balance}) is the relativistic version of the ``ponderomotive force'' exerted by the laser pulse, which is a fundamental concept in laser-driven plasma-based electron acceleration \cite{opticallaser}. See ref. \cite{ponderomotive} for a derivation of the relativistic ponderomotive force using the method of multiple scales.

The conservation of wave action, equation (\ref{wave_action_conservation}), emerges when the vector potential of the laser pulse is subject to the eikonal approximation \cite{envelopeequations}. Hence, the local angular frequency and local wave vector of the laser pulse can be expressed in terms of the phase $\widetilde\Phi$ of the laser pulse as ${\bf k}_0 = \widetilde{\bm{\nabla}}\widetilde\Phi$, $\omega_0 = -\partial_{\tilde t}\widetilde\Phi$; hence the expression
\begin{equation}
\label{local_dispersion_relation_potentials}
\omega^2_p = (\partial_{\tilde{t}}\widetilde\Phi)^2 - c^2(\widetilde{\bm{\nabla}}\widetilde\Phi)^2
\end{equation}
for $\omega_p^2$ follows from (\ref{local_dispersion_relation}).

Expressing (\ref{wave_action_conservation}) and (\ref{number_conservation}) entirely in terms of $\widetilde\Phi$, $\widetilde\Psi$, and fundamental constants, gives the Euler-Lagrange equations for $\widetilde\Phi$ and $\widetilde\Psi$ obtained from the Lagrangian density ${\cal L}=\varepsilon_0 \omega_p^2 \langle {\bf A}_0^2 \rangle/2$. Thus, we have the action
\begin{equation}
\label{dimensionfulAction_3d}
\begin{aligned}
S[\tilde{\Phi},\tilde{\Psi}]=\frac{\varepsilon_0}{e^2c^2}\frac{1}{2}\int d^4\tilde{x}\big\{ ((\partial_{\tilde{t}}\tilde\Phi)^2-c^2(\widetilde{\bm{\nabla}}\tilde\Phi)^2)\\
((\partial_{\tilde{t}}\tilde\Psi)^2-c^2(\widetilde{\bm{\nabla}}\tilde\Psi)^2-c^4)   \big\}.
\end{aligned}
\end{equation}
{\color{black} The implications of (\ref{dimensionfulAction_3d}) can be readily analysed by introducing a dimensional reduction, and the details of the reduction follow from the properties of standard models of laser pulses used in laser-plasma physics (see, for example, ref.~\cite{cernacceleratorschool}). In such models, the pulse envelope is expressed as a product of a Gaussian function of $\tilde{z} - v_g\tilde{t}$, where $v_g$ is the group speed of the pulse, and Gaussian functions of $\tilde{x}$, $\tilde{y}$. The widths of the Gaussians in $\tilde{x}$, $\tilde{y}$ depend on $\tilde{z}$, and are affected by the plasma. Indeed, the interaction between the pulse and plasma is vital as it is used to control the behaviour of the pulse; for example, the diffraction length of a pulse propagating through a suitably pre-formed plasma channel is, by design, many multiples of the vacuum diffraction (i.e. Rayleigh) length~\cite{cernacceleratorschool, opticallaser}. Therefore, we introduce the approximation
\begin{equation}
\int d\tilde{x} d\tilde{y}\,\omega_p^2\langle {\bf A}_0^2 \rangle \approx \tilde{\Lambda} \{\omega_p^2 \langle \textbf{A}_0^2 \rangle\}|_{\tilde{x}=\tilde{y}=0}
\end{equation}
where, in general, $\tilde{\Lambda}$ depends on $(\tilde{z},\tilde{t})$ and is given by 
\begin{equation}
\tilde{\Lambda}=\frac{\text{max}\{\omega_p^2 \langle \textbf{A}_0^2 \rangle\}_{\tilde{x},\tilde{y}}\int_\mathcal{S} d\tilde{x}d\tilde{y}}{\{\omega_p^2 \langle \textbf{A}_0^2 \rangle\}|_{\tilde{x}=\tilde{y}=0}}.
\end{equation}
The domain $\mathcal{S}$ is the region on which $\omega_p^2 \langle \textbf{A}_0^2 \rangle$ is non-zero, and $\int_\mathcal{S} d\tilde{x}d\tilde{y}$ is the cross-sectional area of the laser pulse (spot area). The quantity ${\rm max} \{\dots\}_{\tilde{x},\tilde{y}}$ is the maximum value of the bracketed function in $(\tilde{x},\tilde{y})$, and the line $\tilde{x} = \tilde{y} = 0$ lies along the centre of the laser pulse. Hence, the action (\ref{dimensionfulAction_3d}) has the approximate form
\begin{equation}
\label{dimensionfulAction}
\begin{aligned}
S[\tilde{\Phi},\tilde{\Psi}]\approx \frac{\varepsilon_0}{q^2c^2}\frac{1}{2}\int d\tilde{t} d\tilde{z}\tilde{\Lambda}\big\{ ((\partial_{\tilde{t}}\tilde\Phi)^2-c^2(\partial_{\tilde{z}}\tilde\Phi)^2)\\
((\partial_{\tilde{t}}\tilde\Psi)^2-c^2(\partial_{\tilde{z}}\tilde\Psi)^2-m^2c^4)   \big\}\vert_{\tilde{x}=\tilde{y}=0},
\end{aligned}
\end{equation}
when the derivatives in the $\tilde{x}$, $\tilde{y}$ plane are negligible.}

The above considerations suggest a relativistic bi-scalar field theory given by the action
\begin{equation}
\label{dimensionlessAction}
S[\Phi,\Psi] = \frac{1}{2}\hbar \int d^2x \sqrt{-\eta} \Lambda \eta^{\mu\nu} \partial_\mu\Phi \partial_\nu\Phi (\eta^{\sigma\tau} \partial_\sigma\Psi \partial_\tau\Psi +1),
\end{equation}
where $\eta_{\mu\nu}$ is the background Minkowski metric with signature $(-,+)$, and $\mu,\nu=0,1$. The fields $\Phi$, $\Psi$, $\Lambda$, $\eta_{\mu\nu}$ and the coordinates $x^0=t$, $x^1=z$ are dimensionless.  The action (\ref{dimensionfulAction}) is obtained from (\ref{dimensionlessAction}) using the substitutions
\begin{equation}
\begin{aligned}
& t=\frac{c\tilde{t}}{l_*}, \,\,\,
z=\frac{\tilde{z}}{l_*}, \,\,\,
\tilde{\Lambda}=l_*^2 \Lambda \\
\tilde{\Phi}\vert_{\tilde{x}=\tilde{y}=0} & = \sqrt{ \frac{\hbar e^2}{\varepsilon_0 m_e^2 c^3 l_*^2} }\Phi, \,\,\,
\tilde{\Psi}\vert_{\tilde{x}=\tilde{y}=0}= m_ecl_*\Psi,
\end{aligned}
\end{equation}
where the length scale $l_*$ has been introduced to facilitate the non-dimensionalisation and has no direct physical significance. The field equations arising from the variation of the action (\ref{dimensionlessAction}) are
\begin{equation}
\label{fieldEq1}
\partial_\mu(\Lambda\partial_\nu\Phi\partial^\nu\Phi\partial^\mu\Psi)=0,
\end{equation}
\begin{equation}
\label{fieldEq2}
\partial_\mu (\Lambda(\partial_\nu \Psi \partial^\nu \Psi +1) \partial^\mu \Phi)=0,
\end{equation}
where indices are raised using the background metric. Eqs. (\ref{fieldEq1}), (\ref{fieldEq2}) allow us to derive two effective metrics through a linearisation process.

\section{Effective metric derivation}
Consider the perturbed fields $\Psi=\Psi_0+\epsilon \Psi_1+ \mathcal{O}(\epsilon^2)$, $\Phi=\Phi_0+\epsilon \Phi_1 + \mathcal{O}(\epsilon^2)$, where $\epsilon$ is the perturbation parameter and $\Phi_0$, $\Psi_0$ solve eqs. (\ref{fieldEq1}), (\ref{fieldEq2}) exactly. Field eqs. (\ref{fieldEq1}), (\ref{fieldEq2}) in first order of $\epsilon$ give
\begin{equation}
\label{linearised_fieldEq1}
\partial_\mu(2\Lambda \partial_\nu \Phi_0 \partial^\nu \Phi_1 \partial^\mu \Psi_0 + \Lambda \partial_\nu \Phi_0 \partial^\nu \Phi_0 \partial^\mu \Psi_1)=0,
\end{equation}
\begin{equation}
\label{linearised_fieldEq2}
\partial_\mu (2\Lambda(\partial_\nu \Psi_0 \partial^\nu \Psi_1) \partial^\mu \Phi_0 + \Lambda(\partial_\nu \Psi_0 \partial^\nu \Psi_0+1) \partial^\mu \Phi_1)=0,
\end{equation} 
respectively. The perturbations $\Phi_1$, $\Psi_1$ are coupled and, in general, their field equations (\ref{linearised_fieldEq1}), (\ref{linearised_fieldEq2}) cannot be readily expressed in a manner that reveals one or more effective metrics. However, a pair of effective metrics follows from considering high frequency perturbations of the form $\Psi_1 = {\rm Re} ( a \exp(iK/\check{\epsilon}))$, $\Phi_1 = {\rm Re} ( b \exp(iK/\check{\epsilon}))$, where $\check{\epsilon}$ is a parameter that facilitates the approximation. In this case, the lowest order of $\check{\epsilon}$ yields
\begin{equation}
\Lambda\partial_\mu K[2b \partial^\mu \Psi_0 \partial_\nu \Phi_0 \partial^\nu K + a \partial^\mu K \partial_\nu \Phi_0 \partial^\nu \Phi_0]=0.
\end{equation}
\begin{equation}
\Lambda \partial_\mu K [2a\partial^\mu \Phi_0 \partial_\nu \Psi_0 \partial^\nu K + b\partial^\mu K (\partial_\nu\Psi_0 \partial^\nu \Psi_0 +1)]=0.
\end{equation}
These equations can be written as the matrix
\begin{equation}
\label{matrix}
\begin{pmatrix}
\Lambda \partial_\mu K \partial^\mu K \partial_\nu \Phi_0 \partial^\nu \Phi_0 
& 2 \Lambda \partial_\mu K \partial^\mu \Psi_0 \partial_\nu K \partial^\nu \Phi_0
\\
2 \Lambda \partial_\mu K \partial^\mu \Psi_0 \partial_\nu K \partial^\nu \Phi_0 
& \Lambda \partial_\mu K \partial^\mu K (\partial_\nu \Psi_0 \partial^\nu \Psi_0 +1)
\end{pmatrix}
\end{equation}
acting on $(a, b)^{\rm T}$. The determinant of the matrix in (\ref{matrix}) must be zero so that $a$, $b$ are non-zero. The determinant can be factorised to give two effective metrics $g^{\rm eff}_{\mu\nu,+}$, $g^{\rm eff}_{\mu\nu,{\color{black}-}}$ whose inverses are
\begin{equation}
\begin{aligned}
g^{\mu\nu}_{\rm eff,+}=s \Lambda \{ & \eta^{\mu\sigma} \eta^{\nu\tau}(\partial_\sigma\Phi_0 \partial_\tau\Psi_0 + \partial_\tau \Phi_0 \partial_\sigma \Psi_0)  +\\
&\sqrt{(\eta^{\sigma\tau} \partial_\sigma \Psi_0 \partial_\tau \Psi_0 +1) \eta^{\gamma\delta} \partial_\gamma \Phi_0 \partial_\delta \Phi_0} \eta^{\mu\nu} \},
\end{aligned}
\end{equation}
\begin{equation}
\begin{aligned}
g^{\mu\nu}_{\rm eff,-}=s \Lambda \{ & \eta^{\mu\sigma} \eta^{\nu\tau}(\partial_\sigma\Phi_0 \partial_\tau\Psi_0 + \partial_\tau \Phi_0 \partial_\sigma \Psi_0)  -\\
&\sqrt{(\eta^{\sigma\tau} \partial_\sigma \Psi_0 \partial_\tau \Psi_0 +1) \eta^{\gamma\delta} \partial_\gamma \Phi_0 \partial_\delta \Phi_0} \eta^{\mu\nu} \},
\end{aligned}
\end{equation}
where the constant $s$ satisfies $s^2 =1$ and has been introduced for convenience later in the analysis. It is straightforward to confirm that the determinant of the matrix (\ref{matrix}) can be written as $g^{\mu\nu}_{\rm eff,+}K_\mu K_\nu g^{\sigma\tau}_{\rm eff,-}K_\sigma K_\tau$, where $K_\mu=\partial_\mu K$. Note that raising and lowering indices is done with the background metric $\eta^{\mu\nu}$. The properties of the effective metrics depend on the properties of the fields $\Phi_0$, $\Psi_0$. We will present a regime in which one of these effective metrics is conformally related to the Schwarzschild metric.

\section{Analogue Schwarzschild spacetime}
{\color{black} It will now be shown that the $\tilde{z}$ coordinate introduced in section~\ref{sec:laser-driven_plasma} can be associated with the radial coordinate of a spherically symmetric spacetime. In particular, to obtain an effective metric that is conformally related to the exterior Schwarzschild metric, the ratio of the diagonalised effective metric components must be identified with the ratio of the Schwarzschild metric components $(1-2GM/(c^2r))$ and $(1-2GM/(c^2r))^{-1}$. 

The spot area of the laser pulse is commonly expressed as a function of $z$ only~\cite{opticallaser} and, in practice, the local plasma density is a controllable function of $z$. Hence, it is admissible to choose $\Lambda$ to be a function of $z$ only. Furthermore, we will focus on the properties of a long laser pulse whose longitudinal envelope is substantially larger than the distance over which the changes in the local plasma density are appreciable. Hence, for simplicity, we neglect the dependence of the unperturbed longitudinal envelope on the phase of the laser pulse. The unperturbed, scaled, phase of the laser pulse and the unperturbed dimensionless momentum potential are then
of the form $\Phi_0= \gamma_\Phi t +h_\Phi(z)$, $\Psi_0 = \gamma_\Psi t + h_\Psi(z)$ for some constants $\gamma_\Phi$, $\gamma_\Psi$ and some functions $h_\Phi$, $h_\Psi$. The properties of $h_\Phi$, $h_\Psi$ are determined from those of the desired effective spacetime geometries, and the corresponding local plasma density and local laser intensity can then be constructed using (\ref{average_A0_squared}), (\ref{local_momentum_balance}).}

 Introducing two transformations $\tau_\pm=a_\pm t\pm\mathfrak{f}_\pm(z)$ for some constants $a_\pm$, 
and choosing $\mathfrak{f}_\pm(z)$ such that the effective metrics become diagonal, gives the requirement
\begin{equation}
\label{s.long}
-\left( 1-\frac{z_S}{z}\right)^2 =
\frac{a_\pm^{-2}\left(2h_\Phi'h_\Psi'\pm\sqrt{(h_\Phi^{\prime 2}-\gamma_\Phi^2)(h_\Psi^{\prime 2}-\gamma_\Psi^2+1)}\right)}{2\gamma_\Phi\gamma_\Psi\mp\sqrt{(h_\Phi^{\prime 2}-\gamma_\Phi^2)(h_\Psi^{\prime 2}-\gamma_\Psi^2+1)}-\frac{(\gamma_\Psi h_\Phi'+\gamma_\Phi h_\Psi')^2}{2h_\Phi'h_\Psi'\pm\sqrt{(h_\Phi^{\prime 2}-\gamma_\Phi^2)(h_\Psi^{\prime 2}-\gamma_\Psi^2+1)}}}
\end{equation}
where the dimensionless quantity $z_S$ corresponds to the horizon in the Schwarzschild metric, prime denotes derivative with respect to $z$, and $\pm$ corresponds to $g_{\rm eff, \pm}^{\mu\nu}$. Eq (\ref{s.long}) suggests using scaled variables $h_\Phi=\gamma_\Phi\check{h}_\Phi$, $h_\Psi=\gamma_\Psi\check{h}_\Psi$. Now eqs. (\ref{fieldEq1}), (\ref{fieldEq2}) yield the relationship
\begin{equation}
\label{fieldEqRelation}
\beta (\varepsilon^2-(\check{h}_\Psi')^2)\check{h}_\Phi'=(1-(\check{h}_\Phi')^2)\check{h}_\Psi',
\end{equation}
where $\beta$ is a constant of integration and $\varepsilon^2 =(\gamma_\Psi^2-1)/\gamma_\Psi^2$. Note that
\begin{equation}
\label{constraints}
(\check{h}_\Psi')^2<\varepsilon^2<1
\end{equation}
is required, where the upper bound comes from the definition of $\varepsilon$, while the lower bound is required so that the effective metric components do not become imaginary. Introducing the field $h = \check{h}_\Psi'/\varepsilon$ and choosing $\beta = 1/\varepsilon$ results in $\check{h}_\Phi'=h$ from eq. (\ref{fieldEqRelation}). Furthermore $h^2<1$ is required by (\ref{constraints}). It is now instructive to consider the individual components of the inverses of the effective metrics in $t$, $z$ coordinates which are given by
\begin{equation}
\label{g00}
g_{\rm eff,\pm}^{00}= s\Lambda\gamma_\Phi\gamma_\Psi(2 \mp |\varepsilon|(1-h^2)),
\end{equation}
\begin{equation}
\label{g11}
g_{\rm eff,\pm}^{11}= s\Lambda\gamma_\Phi\gamma_\Psi(2\varepsilon h^2\pm|\varepsilon|(1-h^2)),
\end{equation}
\begin{equation}
\label{g10}
g_{\rm eff,\pm}^{10}= -s\Lambda\gamma_\Phi\gamma_\Psi(1+\varepsilon)h.
\end{equation}
Note that by definition $\Lambda>0$, thus the choice $s=-1$ is required in order to match the signatures of the effective metrics to the background metric in regions where both effective metrics are Lorentzian. Furthermore the $g_{\rm eff,\pm}^{00}$ component and the off-diagonal terms, as well as $\Lambda$, will be non-zero for all values within the constraints (\ref{constraints}). However equating $g_{\rm eff,\pm}^{11}$ to zero and solving for $h$ leads to a horizon if $\varepsilon$ is chosen appropriately. The value of $h$ when (\ref{g11}) equals zero satisfies $h^2=|\varepsilon|/(|\varepsilon|\mp2\varepsilon)$. When $\varepsilon$ is positive, there will be no horizon in $g^{\mu\nu}_{\rm eff,+}$, while there will be a horizon at $h^2=1/3$ in $g^{\mu\nu}_{\rm eff,-}$. The converse is true for $\varepsilon<0$. Since these two outcomes are equivalent, the case of $\varepsilon<0$ will be assumed henceforth without loss of generality. Also note that it can be shown that $\det(g_{\rm eff,\pm}^{\mu\nu})<0$ for $-1<\varepsilon<0$ and $1/3<h^2<1$, thus both effective metrics are Lorentzian. Only $g_{\rm eff,+}^{\mu\nu}$ is of interest and will be explored further, as the other effective metric does not contain a horizon. Let $\nu=-\varepsilon$ for convenience. All of the above considerations let us write the right-hand side of eq. (\ref{s.long}) as 
\begin{equation}
\label{RHS}
\frac{a_+^{-2}\left(1-3h^2\right)^2\nu^2}{\nu(1-3h^2)(2-\nu(1-h^2)) -(1-\nu)^2h^2}.
\end{equation}
Note that the numerator is always positive, and the denominator is proportional to the determinant of $g_{\rm eff, +}^{\mu\nu}$ and as such (\ref{RHS}) is always negative for $1/3<h^2<1$ and $0<\nu<1$. Thus eq. (\ref{s.long}) will always have a solution for $h(z)$ in the specified range. An expression for $a_+$ is obtained by matching the limit of $z\rightarrow\infty$ to $h\rightarrow 1$, yielding
\begin{equation}
\label{aPlus}
a_+^2=\left(\frac{2\nu}{1+\nu}\right)^2.
\end{equation}
An algebraic solution to eq. (\ref{s.long}) can be found since it is a quadratic equation in $h^2$, however the solution is cumbersome and a simpler approach is available for establishing the behaviour of the laser-driven plasma. By taking the square root of eq. (\ref{s.long}), differentiating with respect to $z$ and evaluating at $z=z_S$ gives $h'\vert_S = -a_+(1-\nu)/(6\nu z_S)$, where $\vert_S$ indicates evaluation at $z=z_S$, and $h\vert_S=1/\sqrt{3}$ has been used. The constant $a_+$ must be negative because $h^2>1/3$, $h'\vert_S>0$, thus the negative root of eq. (\ref{aPlus}) is required, and hence
\begin{equation}
h'\vert_S = \frac{1}{3}\frac{1-\nu}{1+\nu}\frac{1}{z_S}.
\end{equation}
Introducing the dimensionless amplitude of the laser pulse $a_0$ given by $a_0=e\sqrt{\langle {\bf A}_0^2\rangle}/(m_ec)$ and using eq. (\ref{average_A0_squared}) gives an expression for $\nu$:
\begin{equation}
\nu=\left.\frac{\sqrt{3}a_0}{\sqrt{2+3a_0^2}}\right\vert_S.
\end{equation}
Note that $\Lambda\propto [h(1-h^2)]^{-1}$ follows from eqs. (\ref{fieldEq1}), (\ref{fieldEq2}). It follows that $\Lambda'\vert_S=0$ and it is straightforward to show that $(\Lambda''/\Lambda)\vert_S=9h^{\prime 2}\vert_S$, which can be used to obtain
\begin{equation}
\tilde{z}_S=\frac{1-\nu}{1+\nu} \left.\sqrt{\frac{\tilde{\Lambda}}{d^2\tilde{\Lambda}/d\tilde{z}^2}}\right\vert_S.
\end{equation}

The effective metric $g_{\rm eff,+}^{\mu\nu}$ is conformally related to the Schwarzschild metric. However, the conformal factor is regular at the event horizon, and the surface gravity and Hawking temperature are independent of this conformal factor \cite{HawkingConformalInvariance}. By construction $\tilde{z}_S=GM/c^2$ and thus, using $T_H=\hbar c^3/(8\pi k_B GM)$, the analogue Hawking temperature is given by
\begin{equation}
T_H=\frac{\hbar c}{8\pi k_B}\frac{1+\nu}{1-\nu}\left.\sqrt{\frac{d^2\tilde{\Lambda}/d\tilde{z}^2}{\tilde{\Lambda}}}\right\vert_S.
\end{equation}
The Hawking temperature can be calculated if the dimensionless amplitude $a_0$ and the laser cross-sectional area $\tilde{\Lambda}$ near $\tilde{z}=\tilde{z}_S$ are known. Since $\Lambda'\vert_S=0$, $\tilde{\Lambda}$ can be expressed as
\begin{equation}
\tilde{\Lambda}=\tilde{\Lambda}\vert_S\left( 1 + \frac{1}{2}\frac{(\tilde{z}-\tilde{z}_S)^2}{\tilde{l}^2} +
\mathcal{O}\left( (\tilde{z}-\tilde{z}_S)^3 \right) \right),
\end{equation}
where $\tilde{l}$ has dimensions of length. It follows that $(\tilde{\Lambda}/(d^2\tilde{\Lambda}/d\tilde{z}^2))\vert_S=\tilde{l}^2$, and thus the Hawking temperature is given by
\begin{equation}
T_H=\frac{\hbar c}{8\pi k_B}\frac{1+\nu}{1-\nu}\frac{1}{\tilde{l}}.
\end{equation}
For practical reasons, $\tilde{l}$ cannot be less than approximately the plasma wavelength $\lambda_p$, and the dimensionless amplitude should satisfy $a_0\leq 1$. As an example $\lambda_p\approx 30$ $\mu$m is achievable \cite{opticallaser} for maintaining an intense near-IR laser pulse propagating through a plasma. This results in $\nu\lesssim 0.77$, the mass of the effective black hole satisfies $M\gtrsim 5.1\times 10^{21}$ kg, and the associated Hawking temperature satisfies $T_H\lesssim 25$ K.

\section{Conclusion}
We obtained the analogue Hawking temperature of a laser-driven plasma system. Perturbed field equations governing a laser-driven plasma were linearised, and the perturbations were assumed to have high frequency in order to derive two effective metrics. The required properties of the fields were found such that one of the effective metrics is conformally related to the Schwarzschild metric. An expression for the Hawking temperature associated with the analogue black hole has been derived. This temperature depends on the values of the dimensionless amplitude and the laser spot area near the analogue event horizon. We have presented one possible way of determining the spot area in terms of the plasma wavelength, with which we demonstrated that for a high-intensity near-IR laser the analogue Hawking temperature is less than approximately $25$ K.

In common with standard analytical treatments of laser-driven plasma accelerators, our results are based on a `cold', collisionless, model of the plasma electrons. However, a comparison of our results and typical plasma temperatures suggests that a detailed model of the laser-driven plasma is needed to confidently identify signatures of the analogue Hawking effect. The temperature of the plasma electrons in a laser-driven plasma accelerator is $\sim 5\times 10^5$ K \cite{opticallaser}, which is $\sim 2\times 10^4$ times larger than the expected analogue Hawking temperature.

Even so, for comparison, it is claimed \cite{BECresult} that an analogue Hawking temperature of $1.2$ nK has been measured in an atomic Bose-Einstein condensate, although these results are disputed \cite{BECdoubt}. Whilst it is clear that identifying the analogue Hawking effect in a laser-driven plasma accelerator is a significant challenge,
the fact that our results show that its analogue Hawking temperature is ten orders of magnitude larger than that of a Bose-Einstein condensate suggests that further investigation is deserved.

\section{Acknowledgements}
This work was supported by the UK Engineering and Physical Sciences Research Council grant EP/N028694/1 (D.A.B.), and the Lancaster University Faculty of Science and Technology (C.F.). All of the results can be fully reproduced using the methods described in the article.

\end{document}